\documentstyle[12pt]{article}
\begin{document}
\vspace*{-2mm}
\thispagestyle{empty}
\noindent
\hbox to \hsize{
\hskip.5in \raise.1in\hbox{\bf University of Wisconsin - Madison}
\hfill$\vcenter{\hbox{\bf MAD/PH/844}
                \hbox{\bf RAL-94-070}
                \hbox{\mbox{}}
            }$
               }
\mbox{}
\hfill July  1994   \\   
\vspace{0.5cm}
\begin{center}
  \begin{Large}
  \begin{bf}
 CALCULATION OF Z PLUS FOUR JET\\ PRODUCTION AT THE TEVATRON\\
  \end{bf}
  \end{Large}
  \vspace{0.8cm}
V. Barger and E. Mirkes\\[2mm]
{\it Physics Department, University of Wisconsin, Madison,
WI 53706, USA }\\
\vspace{0.3cm}
\vspace*{0.3cm}
R.J.N.~Phillips\\[2mm]
{\it Rutherford Appleton Laboratory, Chilton, Didcot, Oxon OX11 0QX, UK
     }\\
\vspace{0.3cm}
and\\
\vspace*{0.3cm}
T.~Stelzer\\[2mm]
{\it Physics Department, Durham University, Durham DH1 3LE, UK.
     }\\[2cm]
  {\bf Abstract}
\end{center}
\begin{quotation}
\noindent
We present the first calculation of $Z+4$ jet production with heavy
quark flavor identification at the Tevatron $p\bar p$ collider.
The $Z + 4$ jet channel is especially interesting as a normalizer for the
$W +4$ jet background to top quark signals, as a background to a possible
$t\to cZ$ flavor-changing neutral-current (FCNC) decay signal, and as a
background to missing-$p_T$ signals from gluino pairs.
We also calculate the  contributions to $W+4$ jet production
from all the different heavy-flavor final states.
The MADGRAPH program is used to generate all leading order
subprocess helicity amplitudes. We  present Monte Carlo results
with  separation and acceptance criteria suitable for the
Tevatron experimental analyses.
The dependence of the cross sections on experimental cuts and the
theoretical ambiguities due to the scale dependence are discussed.
The predicted ($W+4$ jet)/($Z+4$ jet) ratio is insensitive to
most of these choices.
\end{quotation}
\newpage
There are many potential new physics processes at hadron colliders, that
would lead to final states with a weak boson
plus multi-jets, where the weak boson is identified by its
leptonic decay; these signals sometimes also contain a
second weak boson, whose hadronic decay is less easily identified.
Since a weak boson can also be produced along with gluon and quark jets,
a knowledge of these QCD backgrounds is essential to the identification
of new physics signals.  Considerable effort has been devoted in recent years
to the calculation of QCD $W + n$ jet ($n=1,2,3,4$) and $Z + n$ jet
($n=1,2,3$) cross sections; for the cases of high
jet multiplicities $n$, that would be given by many
interesting new physics signals, these calculations
can currently be made at tree level only \cite{vecbos,hag,wisc,mangano}.
We present first results for $Z$ production with
four QCD jets, evaluated for the Tevatron $p\bar p$ collider at
$\sqrt s=1.8$~TeV, including a separation of contributions from
different heavy quark flavors.  We also calculate $W + 4$ jets
with heavy quark flavor identification; this goes beyond previous
$W + 4$ jet results that flag only $b$-flavor \cite{vecbos}.

Major areas of physics interest in a QCD $Z + 4$ jet calculation
are the following.

\noindent
(a) The most immediate interest is related to the
top-quark search at the Tevatron \cite{cdf,d0}, where in the single-lepton
signal with a $b$-tag
the QCD $W+4$ jet channel gives the major background, and a comparison
of the $W/Z$ ratio could provide a calibration; this ratio should be
insensitive to theoretical uncertainties
in the individual cross sections.
Furthermore, experimental acceptance and detector effects  are also
expected to cancel in the ratio.
By calculating separate
cross sections for different final-state quark flavors, we are able to
apply our results to the case where a heavy quark is tagged.

\noindent
(b) Possible isosinglet heavy quarks $x$ would have both charged-current
and neutral-current decay modes, with branching fraction
ratios \cite{barg,bargp,rizzo,aguila,ma}
\begin{equation}
B(x\to qW)\; :\; B(x\to q'Z) \simeq 2\; : \; 1 .
\end{equation}

\noindent
(c) A related question is the possible existence of a prominent FCNC
decay mode of the top quark \cite{hall,agra,mukho,cheng},
$t\to cZ$ along with the standard
$t\to bW$ decay.  In this scenario, $t\bar t$ production would lead to a
$t\bar t\to (cZ)(bW) \to Z + 4$ jet signal, that must be distinguished from
QCD background.

\noindent
(d) The production of supersymmetric particles gives rise to missing-$p_T$
plus multijet signals at hadron colliders.  In particular, production
of gluino pairs $\tilde g \tilde g$  with decays $\tilde g \to \chi_1^0
q \bar q$ to the lightest neutralino $\chi_1^0$ are expected to be a
source of missing-$p_T$ plus 4 jets.    Here $Z + 4$ jet production
with invisible $Z\to\nu\bar\nu$ decays is the dominant standard physics
background.  In the case of $b$-tagged events,
there are regions of parameter space where $\tilde g\to t\tilde t$
or $\tilde g\to b\tilde b$  decays are dominant \cite{baer}.


We now turn to the method used in our $Z + 4$ jet calculation.
An impediment in calculating subprocesses with many final partons is the
large number of Feynman diagrams to be enumerated and expressed as
amplitudes.  For example, the $gg \to Zq\bar q gg$ subprocess involves
516 diagrams.  This phase of the calculation can be accomplished  for
any given subprocess by the MADGRAPH program \cite{madgraph}, which
automatically generates all Feynman graphs and their helicity amplitudes,
employing the HELAS approach \cite{helas}.
However, MADGRAPH does not enumerate
the contributing subprocesses, which must be entered individually, nor
does it carry through the cross section calculation, folding in initial
parton distributions and final phase space integration.  We have added
a phase-space generator and folded in the MRS set $D_{-}^{\prime}$
 parton distributions
\cite{mrs}, evaluated at a scale $Q^2= \left<p_T\right>^2+M_Z^2$,
where $\left<p_T\right>$ is the average transverse momentum of the partons.
The renormalization scale in $\alpha_s$ is set equal to $Q^2$
and the $\Lambda$ value is chosen accordingly to the value
in the parton distribution functions with five flavors.
A similar procedure is followed in our $W + 4$ jet calculations.

For semi-realistic simulations, we make parton-level calculations of
$p\bar p\to W(Z)+4$ jets at $\sqrt s = 1.8$ TeV. We identify final
partons with jets when
\vspace{-1mm}
\begin{equation}
p_T(j) > 20\,\mbox{GeV},\,\, |\eta(j)| < 2, \,\,\Delta R(jj) > 0.4,
\end{equation}
\vspace{-1mm}
where $[\Delta R(jj)]^2 = [\Delta\eta (jj)]^2 + [\Delta\phi(jj)]^2$
defines the angular separation between two jets.  A correction must be
made in comparing the parton transverse momentum $p_T$ with the
observed (uncorrected) jet transverse energy $E_T$; according to CDF
simulations \cite{cdf}, typically 5 GeV or more must be added to the latter.
A full simulation including fragmentation and detector characteristics
must be made for detailed comparisons with experiment.

For the case that $Z$ is detected by $Z\to e\bar e$ and $W$ is detected
by $W\to e\nu$, we take the electron and missing transverse momentum
$p\llap/_T^{}$ acceptance to be
\vspace{-1mm}
\begin{eqnarray}
 && p_T(e) > 20\rm\ GeV, \quad |\eta(e)| < 1,  \\
 && p\llap/_T^{} > 20\hbox{ GeV\quad (for $W$ events)},
\end{eqnarray}
\vspace{-1mm}
and require that the electrons are isolated from jets
by $\Delta R(ej) > 0.4$.
These acceptance criteria approximate but do not exactly duplicate those
used in Tevatron experimental analyses.

Unless otherwise stated, in  the
following $Z$ denotes $Z \to e^+ e^-$ and $W$ denotes
$W^{\pm}\to e^{\pm} \nu$ ; with these leptonic branching fractions included,
the cross sections times branching fractions are denoted $B\sigma$.
Comparison with experiment requires the inclusion of instrumental
efficiencies also.

The total cross sections with these acceptance criteria are
\vspace{-1mm}
\begin{equation}
B\sigma(Z + 4 \mbox{jet}) = 20.5 \,\, \mbox{fb},\,\,\,\,
B\sigma(W + 4 \mbox{jet}) = 337 \,\, \mbox{fb}.
\end{equation}
\vspace{-1mm}
These predictions for the absolute cross sections are however
sensitive to the choice of the scale $Q^2$ in $\alpha_s$.
For $Q^2=\left<p_T\right>^2$ the cross sections are higher by a factor
2.02 (1.93) for $Z+4$ jet ($W+4$ jet) production.

The relative numerical  contributions to the total cross section
from different subprocesses according to  the number
of quarks involved in the process are (in percentages):
\vspace{-1mm}
\begin{equation}
\begin{array}{lccc}
                 & 2q-4g        & 4q-2g      &  6q  \\
Z+4 \mbox{jet}   &   54.5       &    43.5    &  2   \\
W+4 \mbox{jet}   &   53         &    45      &  2    \>. \\
\end{array}
\end{equation}
\vspace{-1mm}
Thus the six-quark contributions are not very important.  The percentage
contributions from the different initial state configurations are
\vspace{-1mm}
\begin{equation}
\begin{array}{lcccc}
                 &   gg   &  qg     &     gq    &  qq,q\bar q,\bar q\bar q\\
Z+4 \mbox{jet}   &   3    &  20.5   &   20.5    &  56           \\
W+4 \mbox{jet}   &   2.5  &  24     &     24    &  49.5\>.      \\
\end{array}
\end{equation}
\vspace{-1mm}
The gluon-gluon initiated production is rather insignificant at the
Tevatron energy.

Tagging of $b$-quarks is an important means of selecting final states
such as $t\bar t$ and $\tilde g \tilde g$, containing heavy quarks.  In
the CDF top-quark search, two means of tagging are employed, a silicon
vertex detector (SVX) and a soft-lepton tag (SLT); the former identifies
displaced vertices and the latter identifies leptons from $b\to \ell\nu X$
and $b\to c\to \ell\nu X$. However, the situation is complicated by the
possibility that a $c$-jet or a light parton jet may be mistagged as
a $b$-jet.  The probability of tagging any particular final state
therefore depends on the separate probabilities $\epsilon_j$
that any single jet $j=b,c,q/g$  satisfies the tagging criteria.  In our
later assessments of tagging, we will assume the values
$\epsilon_b=0.18$ (e.g. 0.11 from SVX and 0.07 from SLT),
$\epsilon_c=0.05$ and $\epsilon_{q/g}=0.01$, which are approximately
the efficiencies in the CDF top-quark search \cite{cdf}.

For application to tagging studies, we present here
the cross sections in fb for different final flavor configurations.
\vspace{-1mm}
\begin{equation}
\begin{array}{cccll}
b  & c & q/g & B\sigma (Z+4\,\mbox{jets}) & B\sigma (W+4\,\mbox{jets}) \\
4  & - &  -  & 0.002              &   0.05            \\
3  & 1 &  -  & <1\cdot 10^{-3}    &   <1\cdot 10^{-3} \\
3  & - &  1  & 0.003              &   0.007           \\
2  & 2 &  -  & 0.012              &   0.11            \\
2  & 1 &  1  & 0.006              &   0.26            \\
2  & - &  2  & 1.03               &   10.2           \\
1  & 3 &  -  & <1\cdot 10^{-3}    &   <1\cdot 10^{-3} \\
1  & 2 &  1  & 0.003              &   0.007           \\
1  & 1 &  2  & 0.001              &   0.04            \\
1  & - &  3  & 0.14               &   0.65            \\
-  & 4 &  -  & 0.006              &   0.05            \\
-  & 3 &  1  & 0.006              &   0.26            \\
-  & 2 &  2  & 0.88              &    10.2            \\
-  & 1 &  3  & 0.24               &   18.6            \\
-  & - &  4  & 18.1               &   297             \\
\end{array}
\end{equation}
Folding in the $b$-tagging efficiencies given above, we obtain the
following tagged cross sections:
\begin{equation}
\begin{array}{ccc}
\mbox{no. of tags}  &  B\sigma (Z+4\,\mbox{jets}) & B\sigma (W+4\,\mbox{jets})
\\
 \geq  0       &    20.5            &      337.        \\
 \geq  1       &    1.23            &      18.1        \\
 \geq  2       &    0.057           &      0.68        \\
 \geq  3       &    0.001           &      0.01
 \end{array}
\end{equation}
To compare these numbers with
experimental event rates, one has to multiply these cross sections by
efficiency factors for the electrons and muons and also take into account
effects of detector simulations. However, these effects (together
with theoretical uncertainties) are expected to cancel approximately
in the ratio of $(W+4 \mbox{jet})/(Z+4 \mbox{jet})$ cross sections.

The predicted $W/Z$ ratio in 4-jet events with at least one $b$-tag is
about 14.7.
In contrast to the individual cross sections, this ratio is
fairly  insensitive to the choice of the scale $Q^2$ in $\alpha_s$.
With $Q^2=\left<p_T\right>^2$ the ratio is 14.1.
 This number is
also
 fairly insensitive to the jet threshold $p_T$
cut. Even if we relax the $p_T$ and $\eta$ requirements on the fourth
jet (as CDF does to increase statistics in the top-quark sample), this
$W/Z$ ratio remains about 14.  This ratio does however depend on the
lepton rapidity cut; for $|\eta (\ell )| < 2.5$ we obtain a $W/Z$
ratio of about 10.

With 19.2~pb$^{-1}$ luminosity, CDF finds two $b$-tagged $Z+4$ jet events
and seven $b$-tagged $W+4$ jet events, with relaxed
$E_T$ and $\eta$ requirements on the fourth jet.   Although the
statistics are small, this observed  $W/Z$ ratio in 4-jet events
appears to be anomalously low in comparison with the QCD prediction.
If future statistics confirm that the $b$-tagged $W4j/Z4j$ ratio
is indeed significantly lower than the pure QCD ratio, then there must be
new physics in the $Z+4$-jet channel.  Furthermore, if the tagged $W+4$-jet
events are indeed dominated by $t\bar t$ production, as suggested by the
CDF analysis \cite{cdf} and by our results above,
then the tagged $Z+4$-jet events\footnote{
The contribution to the $Z+4$ jet final state from $p\bar{p}\rightarrow
W+Z+2$ jets  \cite{wz2j} with hadronic $W$ decay
is much smaller than the number from the
$Z+4$ jet channel.
}
 are dominated
by new physics beyond the standard model.\\
Interesting possibilities for such new physics include (i) a singlet
charge -1/3 quark $x_b$, that mixes with the $b$-quark and therefore
has a prominent $x_b \to bZ$ decay mode \cite{bargp}, or (ii) FCNC decays of
the top quark $t\to cZ$ that would follow from mixing of $t$ with a
charge 2/3 singlet quark \cite{hall}.  In case (ii), the $b$-tag would
have to be faked by the $c$-jet.

We next consider the QCD $(Z\to\nu\bar\nu)+4$ jet background to the
missing-$p_T$ signals of supersymmetry. We here consider missing-$p_T$
requirements of
\begin{eqnarray}
  p\llap/_T^{} > 50\hbox{ or 100 GeV\quad },
\end{eqnarray}
along with the same jet cuts as before.  The integrated cross sections
are

\begin{equation}
B\sigma (p\llap/_T^{}>50;4\,\mbox{jets})= 283\,\, \mbox{fb}
\hspace{1cm}
B\sigma (p\llap/_T^{}>100;4\,\mbox{jets})=97\,\, \mbox{fb}
\end{equation}
The contribution for different final flavor configurations
are:
\begin{equation}
\begin{array}{cccll}
b  & c & q/g & B\sigma (p\llap/_T^{}>50+4\,\mbox{jets})
             & B\sigma (p\llap/_T^{}>100+4\,\mbox{jets})  \\
4  & - &  -  & 0.021              &  0.005                \\
3  & 1 &  -  & < 1\cdot 10^{-3}   &  < 1\cdot 10^{-3}     \\
3  & - &  1  & 0.025              &  0.005                \\
2  & 2 &  -  & 0.13              &  0.050                \\
2  & 1 &  1  & 0.06               &  0.013                \\
2  & - &  2  & 12.1               &  4.3                  \\
1  & 3 &  -  & < 1\cdot 10^{-3}   &  < 1\cdot 10^{-3}     \\
1  & 2 &  1  & 0.03               &  0.007                \\
1  & 1 &  2  & 0.01               &  0.002                \\
1  & - &  3  & 1.57               &  0.37                 \\
-  & 4 &  -  & 0.07               &  0.025                \\
-  & 3 &  1  & 0.06               &  0.012                \\
-  & 2 &  2  & 10.8               &  3.9                  \\
-  & 1 &  3  & 2.7                &  0.64                 \\
-  & - &  4  & 254.               &  88.5
\end{array}
\end{equation}
Including the tagging efficiences assumed above, the tagged cross
sections are as follows.

\begin{equation}
\begin{array}{ccc}
\mbox{no. of tags}    & B\sigma (p\llap/_T^{}>50+4\,\mbox{jets})
               &        B \sigma (p\llap/_T^{}>100+4\,\mbox{jets}) \\
 \geq  0       &  283.       &    97.              \\
 \geq  1       &  16.1       &    5.56             \\
 \geq  2       &  0.70       &    0.24             \\
 \geq  3       &  0.01       &    0.004
\end{array}
\end{equation}

A detailed consideration of the dynamical distributions of $Z+4$-jet
events will be presented elsewhere.

\begin{flushleft}{\bf Acknowledgments}\end{flushleft}
VB and RJNP thank Peter Landshoff and John Taylor for the hospitality of
DAMTP, Cambridge, where some of this work was done.
We thank C. Campagnari for a helpful correspondence.
This research was supported in part by the U.S.~Department of Energy under
Contract No.~DE-AC02-76ER00881 and in part by the University of Wisconsin
Research Committee with funds granted by the Wisconsin Alumni Research
Foundation.

\bibliographystyle{unsrt}

\end{document}